\documentclass{bmcart}

\usepackage{amsthm,amsmath}
\RequirePackage[numbers]{natbib}
\usepackage[utf8]{inputenc} 
\usepackage{tgheros,txfonts}
\usepackage[T1]{fontenc}
\usepackage{xcolor,graphicx}

\startlocaldefs
\endlocaldefs

\begin{document}

\begin{frontmatter}

\begin{fmbox}
\dochead{Methodology}

\title{Planning a method for covariate adjustment in individually-randomised trials: a practical guide}


\author[
  addressref={aff1,aff2},              
  corref={aff1},                       
  email={tim.morris@ucl.ac.uk}         
]{\inits{T.P.}\fnm{Tim P.} \snm{Morris}}
\author[
  addressref={aff1},
  email={rmjlasw@ucl.ac.uk}
]{\inits{A.S.}\fnm{A. Sarah} \snm{Walker}}
\author[
  addressref={aff2},
  email={Elizabeth.Williamson@lshtm.ac.uk}
]{\inits{E.J.}\fnm{Elizabeth J.} \snm{Williamson}}
\author[
  addressref={aff1},
  email={ian.white@ucl.ac.uk}
]{\inits{I.R.}\fnm{Ian R.} \snm{White}}


\address[id=aff1]{
  \orgname{MRC Clinical Trials Unit at UCL},          
  \city{London},                              
  \cny{UK}                                    
}
\address[id=aff2]{%
  \orgdiv{Department of Medical Statistics},
  \orgname{LSHTM},
  \city{London},
  \cny{UK}
}



\end{fmbox}


\begin{abstractbox}

\begin{abstract} 
\parttitle{Background} 
It has long been advised to account for baseline covariates in the analysis of confirmatory randomised trials, with the main statistical justifications being that this increases power and, when a randomisation scheme balanced covariates, permits a valid estimate of experimental error. There are various methods available to account for covariates but it is not clear how to choose among them.

\parttitle{Methods} 
Taking the perspective of writing a statistical analysis plan, we consider how to choose between the three most promising broad approaches: direct adjustment, standardisation and inverse-probability-of-treatment weighting.


\parttitle{Results} 
The three approaches are similar in being asymptotically efficient, in losing efficiency with mis-specified covariate functions, and in 
handling designed balance. If a marginal estimand is targeted (for example, a risk difference or survival difference), then direct adjustment should be 
avoided because it involves fitting non-standard models that are subject to convergence issues. Convergence is most likely with IPTW. Robust standard errors used by IPTW are anti-conservative at small sample sizes. All approaches can use similar methods to handle missing covariate data. With missing outcome data, each method has its own way to estimate a treatment effect in the all-randomised population. We illustrate some issues in a reanalysis of \textit{GetTested}, a randomised trial designed to assess the effectiveness of an electonic sexually-transmitted-infection testing and results service.


\parttitle{Conclusions} 
No single approach is always best: the choice will depend on the trial context. We encourage trialists to consider all three methods more routinely.
\end{abstract}


\begin{keyword}
    \kwd{Covariate adjustment}
    \kwd{Estimands}
    \kwd{Standardisation}
    \kwd{Inverse probability of treatment weighting}
    \kwd{Randomised controlled trials}
    \kwd{Clinical Trials}
    \kwd{Missing data}
\end{keyword}


\end{abstractbox}
%

\end{frontmatter}



\section*{Background}

Randomised controlled trials are designed to estimate average treatment effects. This article considers the handling of \textit{covariates} in the analysis of individually randomised trials. By \textit{covariates}, we mean measurements on participants recorded at baseline that are thought to be prognostic. Typical examples are age at randomisation and disease severity at randomisation.

Confounding does not affect trials that are properly randomised, since confounding is a systematic bias and any imbalances in covariates in randomised trials are due to chance, which introduces non-systematic error that is reflected in inference\cite{senn13}. However, it remains desirable from a statistical perspective to account for covariates in the analysis of a trial.

Covariate adjustment is desirable because, if a covariate predicts outcome, accounting for its effect on outcome will improve power to detect a treatment effect\cite{kahan14covpow,hernandez06,turner12} unless none of the covariates in a model are prognostic\cite{kahan14covpow}. This is sometimes explained as `accounting for chance imbalance', though we view this differently: a large imbalance (as might be expected in an observational study) can make inference less precise; adjustment gains power by acknowledging chance covariate balance.

Covariate adjustment is also desirable because, if a covariate is balanced by the randomisation scheme, for example by using stratified block randomisation or minimisation, adjustment is necessary to obtain a valid estimate of experimental error. An unadjusted analysis mistakenly assumes that chance imbalance in covariates \textit{could} have occurred -- an extremely useful property for covariates that were unmeasured. However, when randomisation has been balanced according to measured covariates, the fact that imbalance could not have occurred must be acknowledged in our statistical inference\cite{kahan12}. This was pointed out by Fisher almost a century ago but is frequently ignored in practice\cite{fisher1926,kahan12bmj}.

It has further been argued that it is illegitimate to ignore covariates that have been measured; see `myth~6' of reference~\cite{senn13}.

A premise of this article is therefore that accounting for covariates in the analysis of trials is a good idea. We are concerned with how trial statisticians and clinical investigators should agree on a method before seeing the data. This perspective is taken to help statisticians working on trials make sensible, informed decisions when writing a statistical analysis plan.

We will focus most heavily on binary outcome data since this is where some of the issues are most acute.
When the outcome measure is continuous and the analysis aims to estimate a difference in means, some -- though not all -- of the considerations of this paper become redundant. In particular, the discussion below about non-collapsibility is not relevant to the mean difference, which is collapsible (which will be discussed further for binary outcomes). Desirable properties of analysis of covariance using ordinary least squares for continuous outcomes are well appreciated, particularly when treatment--covariate interactions are assumed to be negligible (discussed further below).

For adjustment to be worthwhile, the covariates to be included in the analysis must be prognostic: adjustment for a covariate that is not prognostic is essentially equivalent to an unadjusted analysis, though it can lose power in small samples. It is not the purpose of this article to discuss which covariates to choose. The Committee for Proprietary Medicinal Products \textit{Points to consider on adjustment for baseline covariates} document gives some guidance\cite{cpmpcovariates04}. It does caution against approaches that select covariates most strongly associated with the outcome in the trial; however, subsequent work (for example~\cite{tsiatis08,balzer16}) has shown that such a procedure can be pre-specified in a principled manner.

Part of the motivation for this article is that statistical research papers have frequently recommended covariate adjustment due to improvements in power, with little thought given to the implications of different adjustment methods (TPM being one culpable author). The approach generally favoured in clinical trials is direct adjustment using an outcome regression model. This is so prevalent that reviews of practice have not needed to discuss which methods were actually used\cite{ciolino19,kahan12bmj}. This article works through the implications and aims to contrast its properties with two other methods of adjustment better known in the epidemiology literature: \textit{standardisation} and \textit{inverse probability of treatment weighting}.


\section*{Methods}

\subsection*{Motivating examples: the \textit{GetTested} trial \label{s:gettested}}


The \textit{GetTested} trial was designed to assess the effectiveness of an internet-accessed sexually transmitted infection (STI) testing and results service (chlamydia, gonorrhoea, HIV, and syphilis) on STI testing uptake and STI cases diagnosed.

Briefly, 2,072 participants were recruited in the London boroughs of Southwark and Lambeth. Participants were randomised to an invitation to use an internet-based sexual health service (intervention) or a standard test from a walk-in sexual health clinic (control).

Treatment allocation involved a minimisation procedure balancing for gender (male/female/trans), number of sexual partners in the 12 months before randomisation (categorised as 1, 2+; note that one-or-more was part of the eligibility criteria) and sexual orientation (men who have sex with men vs.\ other groups). Each of the covariates was weighted equally when determining marginal imbalance and intervention or control were assigned at random with 80\% probability of assignment to the favoured arm.

The two outcomes of principal interest were both binary. The first was whether participants took an STI test within six~weeks of randomisation, with the control arm proportion anticipated to be around 10\%. The second outcome, and the outcome for which sample size was calculated, was STI diagnosis (following a positive STI test) within six~weeks of randomisation, with control arm proportion anticipated to be around 0.6\%. The primary analysis was planned to account for the following covariates: gender, age, number of sexual partners in the 12 months before randomisation (10 categories where the final category is $>10$), sexual orientation (four categories) and ethnicity (five categories)\cite{wilson16}. The chosen covariates were all assumed to be prognostic.

\subsection*{Three broad approaches to accounting for covariates} \label{s:3approaches}

Three broad approaches to covariate adjustment are described below. We outline the generic procedures for estimation in this section and discuss their properties in the next section. Note that the estimand targeted is discussed there.

\subsubsection*{Direct adjustment}
`Direct adjustment' refers to fitting an outcome regression model including terms for randomised treatment $Z$ (an indicator equal to 1 if assigned to treatment and 0 if assigned to control) and covariates $X$, but no interaction between $Z$ and $X$. The treatment effect and its standard error are estimated directly as the treatment coefficient from the model. This might be done, for example, using a generalised linear model, which may be standard (with canonical link function) or non-standard,
or a Cox model\cite{McCullaghNelder89,Coxph72}.

\subsubsection*{Standardisation}
Standardisation fits an outcome regression model including $Z$ and $X$, possibly including interactions with $Z$, and then \textit{standardises} the results by summing or integrating over the distribution of covariates observed in the trial. One intuitive way to achieve this is by making predictions if all participants were assigned to the intervention arm and then if all participants were assigned to the control arm and forming a suitable contrast of the predictions\cite{hernan20ci,cummings09,snowden11,lee04}. Standardisation builds on outcome regression by fixing the summary of the treatment effect but allowing flexible modelling to estimate it; for example, a risk ratio can be estimated using logistic regression. Standardisation is sometimes termed \textit{marginalisation} or \textit{G-computation}, and is implemented in Stata's \texttt{margins} command and R's \texttt{stdReg} package\cite{sjolander16}.

\subsubsection*{Inverse probability of treatment weighting}
Inverse probability of treatment weighting (IPTW) involves fitting two models. The first uses participants' covariate values to predict the probability of being randomised to the arm they were in fact randomised to. The second then fits a simple model contrasting the treatment arms, weighted according to the inverse probability of treatment estimated by the first model. This effectively creates a weighted trial sample (pseudo-population) in which both trial arms have the same distribution of observed covariate values. This may sound odd since, if we have used simple randomisation, we \textit{know} that the `true' weight for all participants is identical, so trying to predict it with covariates appears futile\cite{williamson14}. However, the goal is not to obtain an estimate of this probability but instead to either reweight to a better balanced trial or acknowledge the balance observed.

The method is computationally the same as using propensity scores in observational data but the motivation and considerations for variable inclusion are different (just as the motivation for covariate adjustment would be different in randomised trials \textit{vs.}\ observational studies). 

\subsubsection*{Comments on methods}
There are many possible specific ways to implement each of the three broad approaches. For example, Tsiatis and colleagues used a form of standardisation but the within-arm outcome models are defined to yield `as good predictions as possible without concerns over bias'. However, for the purposes of this article, we will generally refer to or use simple implementations of the three broad approaches. For standardisation with binary outcomes we will (by default) use a logistic regression working model fitting main effects of covariates to produce predictions. For IPTW we will use a logistic regression including only the main effects of covariates to model the probability of randomised group given covariates. For direct adjustment, we will use generalised linear models with a link function that permits parameter estimation on the scale of the summary measure of interest, for example a binomial generalised linear model with identity link function for a risk difference.

Besides the three broad approaches there exist hybrid methods; in particular, estimators based on the influence function target marginal summaries and include both a model for treatment (as does IPTW) and a model for outcome (as does standardisation). Further, the superficial relationship between this form of IPTW and propensity scores suggests other estimators, for example matching rather than weighting. However, we regard the possibility of discarding data from some randomised individuals makes it unpalatable and we do not consider it further.




\section*{Results \label{s:properties}}
We now work through some properties that are important to consider when writing a statistical analysis plan. These properties are mainly -- but not solely -- statistical. We also report results of applying the different methods to the \textit{GetTested} trial.

\subsection*{Summary measure of the estimand: marginal or conditional}
The topic of \textit{estimands} has become increasingly prominent in clinical trials since the publication of the ICH~E9(R1) addendum\cite{addendum19}. The addendum lists five attributes of an estimand: the treatment condition of interest, target population, outcome variable, handling of intercurrent events and population-level summary. Here, we focus on the population-level summary (the later section on handling missing data will focus on the target population). Table~\ref{t:psumm} lists some population-level summaries commonly used in clinical trials and, in particular, notes whether conditional and marginal estimands coincide.

It is not our purpose to argue for any particular summary measure but it would be remiss to pass over how the choice should be made. Some statisticians assume that the correct approach must be to choose a measure as a parameter of a model that might have generated the data. Permutt argues that the choice of scale should be `linear in utility': a hypothetical value of treatment effect of~2 should be twice as attractive as 1 whatever the potential outcome on control\cite{permutt20}. Others argue for a measure that can be easily interpreted. We regard the first view as misguided even if the true model were known and the second as too strict, since it rules out any relative measure. The choice of measure should be a trial-specific tradeoff between ease of interpretation, close relation to average patient benefit and potential transportability to other settings.

The following discussion relates to \textit{non-collapsible} summaries\cite{daniel21}.
For readers not familiar with this term, we provide a numerical example in table~\ref{t:collaps} and note . The reader should suppose that the frequencies given are `true' in the sense that, had we recruited a very large sample size, the cells of table~\ref{t:collaps} would contain exact multiples of the frequencies shown.

Consider a trial in a condition that includes participants from two measurable strata, A and B, which have a substantial effect on prognosis. The trial team recruits 40 participants -- 20 from each stratum -- and, within strata, randomises 10 to intervention and 10 to control. In stratum A, the odds of dying on the control arm is $5/5=1$. In stratum B, the outlook is far more favourable, with the odds of dying on the control arm just $1/9$. Despite these differences, the treatment effect (a conditional odds ratio), is 9 in each stratum. We might have conducted a trial recruiting patients from just one of the strata or from both. If we put all 40 people together, as shown at the right-most block of table~\ref{t:collaps} (hence the term \textit{marginal}), the odds of dying in the control arm becomes $(1+5)/(5+9)=3/7$ and the marginal odds ratio is $5.4$.

At first sight this is astonishing! Both strata have odds ratios equal to 9 but, when put together, the odds ratio changes. It is not a weighted average of the within-stratum odds ratios. Treatment is exactly balanced within strata, so this is not due to imbalance. Neither is it effect modification, since the log odds ratio is identical in both strata.

The general phenomenon is known as `non-collapsibility', which describes the relationship between the marginal and conditional summary measure: the true \textit{marginal} odds ratio is attenuated towards~1 compared with the \textit{conditional} odds ratio. It occurs because the average of the logit is not the logit of the average. While the odds ratio is non-collapsible, the risk ratio and risk difference are collapsible. For insights into why non-collapsibility occurs, see Daniel, Zhang and Farewell\cite{daniel21}; see also~\cite{didelez21} and~\cite{huitfeldt19} for more technical discussions, particularly into the relation between collapsibility and confounding.

Rather than collapsing our strata into `both', as in the right-hand panel of table~\ref{t:collaps}, it is possible to adjust for strata and recover the conditional odds ratio of 9 (using logistic regression adjusted for strata, or stratified Mantel--Haenszel).

Note that non-collapsibility is not `bias' as sometimes supposed, but a case of different estimands. The odds ratio of 9 is a within-stratum or \textit{conditional} estimand formed by comparing the effect of treatment within a stratum, relating to the question `what would be the odds ratio for treatment comparing people in the same stratum?' The odds ratio of 5.4 is a between-strata or `marginal' estimand formed by comparing the effect of treatment for groups made up of half stratum A and half B, relating to the question `what would be the odds ratio in a population made up of half stratum~A and half stratum~B'.

Had the notional trial recruited a different proportion of patients from each stratum, the conditional odds ratio would have remained 9 but the marginal odds ratio would not have remained 5.4. For example, suppose that in stratum~A we had recruited three times as many people (so that the numbers in that row are all multiplied by three). The marginal odds ratio would then be 6. Its true value depended on the proportion of participants in each stratum. In general, a marginal summary depends on the covariate distribution when we have non-collapsibility. This is uncomfortable; after all, regression models condition on covariate values rather than modelling their distribution, but we see that the distribution nonetheless matters.

This discomfort may lead us to conclude that the conditional odds ratio is obviously preferable. This is misled for two reasons. First, suppose there were a second covariate to stratify on table~\ref{t:collaps} that also strongly predicted outcome but it was unknown (or at least unmeasured). Then true value of the conditional odds ratio still depends on the distribution of a covariate, but since the covariate is unmeasured its distribution is unknown: we cannot know what we are marginalising over. Second, this example is contrived such that the conditional odds ratios within strata were identical. When there is effect modification on the scale of the population summary measure, the true value of the conditional measure will also depend on the distribution of observed covariates.

Either a marginal or conditional estimand may be desirable and this depends on context. For example, a patient may wish to know `what would happen if someone similar to me were to choose this intervention \textit{vs.}\ not?' Meanwhile, for policy makers, the average difference a drug would make if offered to a group of people might be of more interest, though they might equally wish to know about the effect for specific groups. Note that a different covariate distribution in the target group changes the value of the marginal estimand. Some authors have explored on how to extend inference to a different target population\cite{dahabreh20,cole10}. Interestingly, marginal estimands appear to be favoured for causal inference from observational data: Hern\'an and Robins define a population causal effect as `a contrast of any functional of the \textit{marginal} distributions of counterfactual outcomes under different actions or treatment values' (emphasis added)\cite{hernan20ci}. By this definition, the within-stratum odds ratio of 9 would not target a population causal estimand, since it is a contrast of conditional distributions, though the word \textit{marginal} is a preference rather than necessary to the definition.

In non-inferiority studies, where the null is a non-zero difference between arms on some scale, it is frequently argued that the margin of non-inferiority is easier to understand, define and interpret on a marginal than a conditional scale. This is in line with our own experiences in collaborations.

The choice of marginal or conditional estimand is clearly not simple: the true value of the estimand will depend either on the distribution of observed covariates (always marginal and sometimes conditional), on which covariates are conditioned-on in the model (conditional), and further on the distribution of omitted prognostic covariates (both). Note that these aspects have implications for the quantities being combined in meta-analysis and for apparent heterogeneity in meta-analysis. We will not comment further on these points here.

For non-inferiority studies with a non-collapsible summary measure, it is worth noting the scale on which the non-inferiority margin is defined. Suppose the margin is specified as a marginal hazard ratio, then the corresponding non-inferiority margin on the scale of the conditional hazard ratio is further from 1. If this fact is forgotten and a conditional hazard ratio is estimated without changing the margin, we could expect to lose power compared with an unadjusted analysis (which targets a marginal estimand).

In terms of the three broad methods considered, direct adjustment always targets a conditional summary measure; standardisation typically -- but not necessarily -- targets a marginal summary measure; and IPTW always targets a marginal summary measure. Standardisation and IPTW are rarely used in trials but receive more attention in the epidemiological literature. One justification given for this seems be that the notional `target trial' would always target a marginal summary\cite{hernan20ci}. Ironically, trials which do adjust for covariates tend to use direct adjustment and so target a conditional summary.

\subsection*{Convergence}
Having defined an estimand, we require an estimator to compute an estimate. For many estimators, parameter estimation proceeds through some iterative technique. In maximum likelihood estimation, for example, an algorithm is used to find parameters that maximize the likelihood of the data. This involves finding parameter values that maximize a function. Once an algorithm has found a maximum, it is said to have `converged'. It sometimes happens that the algorithm fails to converge to a maximum or that the maximum to which is converges is local (that is, a small bump rather than the true maximum) or not unique. This is clearly an issue.

In the analysis of randomised trials, non-convergence tends to occur for one of two reasons: first, problems that occur with certain models (e.g.\ generalised linear models with binomial outcome distribution and identity- or log-link function); second, including too many parameters compared with the effective sample size (e.g.\ fitting fixed centre intercepts with few participants per centre)\cite{kim20}. Essentially, the observed data are not consistent with a model that fits within the given constraints.

We are concerned with choosing a procedure for analysis prior to seeing data, so it would be unwise to jeopardise the analysis by choosing a procedure that may not converge\cite{morris14sa,gamble17}. While it may be possible to specify a backup procedure, it would need to target the same estimand. This may prompt the question \textit{why not specify the backup procedure as the first choice?} (one good reason may be due to lower power).

Convergence requires particular attention when inference relies on (for example) bootstrapping or re-randomisation tests. Both involve augmenting the data using simulation and analysing the resulting dataset. We now need to be confident that not only will convergence be achieved in one dataset but in every dataset constructed by the procedure.

Direct adjustment and standardisation may involve fitting different outcome models. For example, to estimate a risk difference, direct adjustment may use a generalised linear model with binomial outcome with identity link function. In terms of convergence, this would be a risky plan. It is possible to estimate an adjusted risk difference using other methods. This could use standardisation following estimation through a logistic regression model, which comes with the guarantee of converging to a unique maximum. A popular technique for estimating the risk ratio without incurring convergence problems is to use a poisson model with robust standard errors\cite{Yelland2011}, where convergence is likely due to the canonical link function.

IPTW involves specifying a model for treatment $P(Z~|~X)$ to estimate weights. This may be any model for binary data (regardless of the trial outcome type). Due to randomisation, it is always true that $P(Z~|~X)=P(Z)$. This means that the model for a binary treatment would not be misspecified regardless of how covariates $X$ are modelled, provided parameters were not \textit{constrained} to be wrong. Allocation ratios in trials are most frequently 1:1 but rarely more extreme than 1:2. For 1:1 allocation, the `outcome' proportion in the treatment model will be approximately 50\%, and never near 0 or 1, and its distribution given covariates is random, making `separation' (or `perfect prediction') unlikely\cite{HeinzeSchemper02}. All this means that the treatment model has a good chance of converging. The subsequent outcome model has no covariates and so is certain to converge.

IPTW therefore seems to be the safest broad approach if convergence is anticipated to be an issue, while standardisation may mitigate possible issues associated with direct adjustment (given the same estimand such as a risk ratio).

\subsection*{Efficiency/precision/power}
A key reason to account for covariates in the analysis is to increase power. Note that for non-collapsible summary measures it is wrong to attempt to compare precision of marginal and conditional estimators but in general it is possible to compare power when the null is zero difference, since collapsibility then holds\cite{daniel21}. Marginal adjusted estimators have been shown to be more efficient than marginal unadjusted estimators\cite{colantuoni15,moore09}.

Because adjustment separates the effect of a treatment from the effects of covariates, we can typically infer the effect of a treatment with a little more precision, though it is possible to lose precision in small samples with non- or weakly-prognostic covariates. It is therefore usually desirable to use an efficient method of accounting for covariate effects, or the potential gains in power may not be fully realised.

While it is sometimes argued that weighting estimators are inefficient, Williamson, White and Forbes showed that, in the trials context with a continuous outcome measure, IPTW is asymptotically as efficient as direct adjustment\cite{williamson14}, backed up by simulation results using finite samples. Any `inefficiency' of IPTW tends to arise due to extreme weights, just as the variance reduction it achieves is a result of similar estimated weights for all individuals. As with convergence, thinking about the weighting model makes clear that this will not be a problem when using this method in the analysis of trials. Note that a closely-related method, overlap weighting, has recently been shown to be more efficient in finite samples and is worth consideration for covariate adjustment\cite{zeng20,desai19}. At the time of writing, the lack of a general implementation in statistical software means, here, we do not further consider this otherwise attractive approach.


\subsection*{Handling covariates balanced by design}
There are many methods of balancing covariates at the design stage. The most popular seem to be stratified blocks and minimisation\cite{kahan12bmj}. When a covariate-balancing method is used, it is necessary to account for the randomisation scheme in the analysis, or the estimated standard error for the treatment effect will be biased upwards, producing confidence intervals that are too wide (meaning they have greater than $1-\alpha$ coverage) and miscalibrated p-values\cite{kahan12}.

For some intuition, suppose a trial uses stratified blocks, with stratification by a single, binary, prognostic covariate. That covariate will then always be distributed equally across the randomised groups (provided each block is completed). It is then impossible for any difference seen to be due to this covariate. Effectively, the variability in the treatment effect due to possible chance imbalances is eliminated, since imbalances can never occur under this design. An analysis that ignores this systematic balance will assume that imbalance in a covariate could have occurred by chance and calculate a standard error accordingly. This would be too large, since an imbalance in this particular covariate could not in fact have occurred\cite{senn13}. Adjustment for the covariate separates the effect of covariate/s on outcome from the effect of treatment on outcome, and this is acknowledged in calculation of the standard error.

There is literature on this going back at least to Fisher, who seemed to regard the point as obvious for analysis of variance in agricultural experiments\cite{fisher1926}. It is generally accepted that direct adjustment and standardisation can be specified to provide valid standard errors. This also holds for IPTW, though has not previously been commented on. Suppose we again have a single binary covariate which is perfectly balanced across treatment groups and estimate $\text{P}(Z=z~|~X=x)$. Then this probability will be identical for every individual in the trial. Fitting the weighted regression to contrast the treatment effect will then return an identical estimate to a model that ignores the covariate. What is perhaps surprising is that the IPTW estimator still has a smaller standard error than the unweighted model. The variance formula, which does not say anything about the design, effectively `sees' and acknowledges the balance after estimating the weights and rewards itself accordingly\cite{williamson14}. It is also clear that this happens by analogy to direct adjustment, since asymptotically the two methods have the same standard errors.

\subsection*{Variance estimation}
Some variance formulas rely on approximations and some are asymptotic. For direct adjustment based on maximum likelihood estimation, formulas are available for all commonly used models. For standardisation, the standard error from a fitted model is transformed using the delta method (asymptotic)\cite{oehlert92}. Due to non-linearity, this could lead to p-values and confidence intervals that do not quite agree, if the p-value is taken from the model on the estimation scale; one possibility is to use test-based confidence intervals for measures with the same null. For example, suppose the outcome model is a logistic regression and the summary measure of the treatment effect is a risk difference. The logistic regression returns a test-statistic of $z$ for treatment. A 95\% confidence interval for the risk difference can then be constructed by taking $\pm 1.96/z$ times the distance between the estimated risk difference and 0, and adding the result to the estimated risk difference.

IPTW estimators use robust standard errors that acknowledge the estimation of weights in the first step. These robust standard errors are asymptotically valid but recent work has demonstrated that they can produce slight undercoverage in small samples\cite{raad20}. In this type of setting, non-parametric bootstrap may be required, until a closed variance formula with a small-sample correction has been developed.

While non-parametric bootstrap is a useful tool, we regard it as not being ideal due to inherent Monte Carlo error -- though it may sometimes be the only option. Monte Carlo error can be made small with a suitably large number of bootstrap repetitions. An important but often neglected point about the bootstrap is that the resampling procedure must mimic the sampling used in the study itself. A simple bootstrap procedure invokes simple randomisation, and will return upwardly biased standard errors or confidence intervals that are too wide if the trial did not in fact use simple randomisation. So, if the trial used blocked randomisation within strata, the bootstrap procedure should be restricted to do the same, else it targets the wrong sampling distribution. Ensuring that the bootstrap procedure mimics the design actually used may be awkward if for example the trial used minimisation.

\subsection*{Misspecification of the covariate model}
We consider misspecification of the mean function relating covariates to outcome rather than misspecification more generally. We illustrate the ideas using a continuous outcome, which lends itself to this visual explanation, but expect similar results for other outcome types.

Consider a study with a single covariate $X_i$, randomised treatment $Z_i$ and the model that generates outcomes $Y_i$ is
\begin{equation} Y_i = \alpha + \theta Z_i + \gamma X_i^2. \end{equation}\label{eq:true}
Note that there is no residual error here; the mean function determines the outcome exactly. The analyst plans to fit to the observed data a model (which is misspecified) with mean function
\begin{equation} y_i = \hat{\alpha} + \hat{\theta} z_i + \hat{\lambda} x_i. \end{equation}\label{eq:ms}
Suppose that this notional study is run and the observed $x$ among those recruited is perfectly uniform on $(-0.5,0.5)$, as depicted by the first horizontal grey bar at the top of figure~\ref{f:quad}. When the misspecified model is fitted, $\hat{\lambda}=0$. Next consider a trial where observed $x_i$ values are uniform on $(0,1)$ or $(0.5,1.5)$, also depicted by grey horizontal bars in figure~\ref{f:quad}. It is now clear that when the analyst fits their model~\ref{eq:ms}, $\hat{\lambda}>0$. In the first case the sample correlation of $x$ with $x^2$ is zero, but in second and third cases it is greater than zero. The analyst's adjustment for $x$ thus partially adjusts for $x^2$ despite the model being misspecified. This will generally be true when a covariate actually adjusted for is correlated with covariates not adjusted for.

The lower panel of figure~\ref{f:quad} gives the estimated standard errors after linear adjustment, showing that linear adjustment is always as efficient as no adjustment.

When using direct adjustment using the data in figure~\ref{f:quad}, the model is misspecified. Meanwhile, when using IPTW, the model used to form weights is by definition correctly specified. Despite this, the two return nearly identical results. The criteria for good specification of IPTW are slightly different than usual: what matters is not the correctness of the specification of the model for $Z~|~X$ but how well the model for $Z~|~X$ models the predictors of $Y$; doing this better will result in a more suitable `rebalancing'. By attempting to balance $X$ instead of $X^2$, the covariate will be well balanced at certain points but less so at (for example) particularly high or low values of $X$.


\subsection*{Handling data missing at random with the three adjustment approaches}\label{s:missing}

Some data will inevitably be missing for some participants in the majority of randomised trials. This is most often in the outcome (unless `missing' is somehow part of the outcome definition) but sometimes occurs in one or more covariates. Meanwhile, data on randomised arm should never be missing. We will first discuss the issues when outcomes are incomplete and then when covariates are incomplete, along with some solutions for each of the three broad approaches. As with any inference from incomplete data, it is important to understand the mechanisms under which bias will and will not be introduced and so we discuss missingness dependent on randomised arm, covariates and outcome separately.

With missing outcome data, a good starting point is to consider implications of missingness for the simplest analysis: including only those individuals with observed outcomes (\textit{complete-case} analysis). When the probability of data being missing depends only on randomised arm, a complete-case estimator is unbiased and efficient. When missingness depends on the outcome, complete-case analysis and multiple imputation under missing-at-random are biased in general; we will return to this point under \textit{sensitivity analysis}.

When outcome missingness depends on the observed value/s of covariates, a complete-case estimator may or may not be biased. If the covariate/s were not adjusted for, data would be missing not at random, which is a more difficult statistical problem; if the covariate/s causing missingness are adjusted for in the analysis, this becomes a missing-at-random problem. It is sometimes said that multiple imputation is not needed when outcomes are missing-at-random. However, estimation based on the complete cases is unbiased for a population represented by the complete cases. This estimand does not in general equal the estimand that targets the population actually randomised, unless the treatment effect is the same for these two populations; potentially a strong assumption. Supposing that the all-randomised population is of interest, a complete-cases estimator is then potentially biased, though the magnitude of bias will be small in practice. For a worked numerical example explaining this point, see the supplementary material.

If the aim of accounting for covariates in the analysis is simply to increase precision or to estimate a conditional summary measure, and not to target the all-randomised population, the remainder of this section and the appendices can be skipped.

With outcomes missing according to this mechanism (depending on covariates), performing multiple imputation by-randomised-arm and then analysing the imputed data sets by any of the three approaches to adjustment targets the all-randomised population\cite{sullivan16}. For direct adjustment, this is straightforward but there are subtleties in terms of statistical inference for the other two approaches:
\begin{itemize}
    \item For standardisation, the question is whether to apply Rubin's rules before or after the standardisation step. Since approximate normality is more likely on the estimation scale (for example log-odds) than the summary scale (for example risk difference), this is likely to be the appropriate scale for combining.
    \item For IPTW, it is however possible that Rubin's variance formula will be inconsistent due to uncongeniality\cite{Meng94}. Further, attempting to use multiple imputation may involve fitting the direct adjustment model -- the first step of standardisation -- so using multiple imputation may imply that IPTW is not needed. We would lose, for example, the  advantages in terms of convergence.
\end{itemize}

There are alternatives to multiple imputation. Under covariate-dependent missingness, standardisation can be applied to the all-randomised sample rather than only those with complete outcome data. 
Meanwhile, IPTW can be combined with inverse probability of missingness weighting, with missingness predicted from covariates separately by randomised arm. These are not our primary recommendations because we want a principal analysis that can be readily extended to principled sensitivity analyses.

Missing covariate values are not inevitable in randomised trials but do sometimes occur. As with missing outcome data, analysis based only on the complete cases may inadvertently target a complete-cases population rather than all randomised, and may be biased if the all-randomised population is intended. Unlike with missing outcome data, discarding individuals with observed data on treatment and outcome does not follow the intention-to-treat principle and does throw away information. However, it is sometimes simple to target the all-randomised population: for any method that targets a marginal or collapsible summary measure, simple mean imputation (across arms, not within) and the missing indicator method are generally appropriate methods\cite{WhiteThompson05}. When using direct adjustment with a non-collapsible summary measure, it is more difficult to deal with incomplete covariate data, and this typically requires a correct model\cite{WhiteThompson05}; if covariates are missing not at random, this can be very difficult. The message is then that more care is required to collect all covariate data if direct adjustment is to be used with a non-collapsible summary measure.

\subsection*{Sensitivity analysis with outcomes missing-not-at-random \label{s:sensitivity}}

There are rarely cases where we \textit{know} the true missingness mechanism. The assumption of missing-at-random depending on randomised arm and covariates is a convenient starting point but it is important to examine the extent to which inferences are sensitive to alternative missingness mechanisms. This prompts \textit{sensitivity analysis}\cite{morris14sa}. To obtain valid inference, missingness mechanisms that represent departures from missing-at-random then need to be explicitly invoked.

In this situation, we view multiple imputation as a general and convenient framework for statistical inference under various departures for each of the three broad approaches, though not the only one\cite{cro20}. Suppose for example that the planned approach to covariate adjustment was standardisation, and under missing-at-random we planned to standardise to the all-randomised sample (which is valid). There is no extension of this concept under missing-not-at-random mechanisms. Multiple imputation by-arm is a convenient way to do sensitivity analysis. It makes little sense to have a mismatch between the primary analysis and sensitivity analyses other than the missingness mechanism invoked, since we want to ensure that sensitivity of results are attributable to the change of missingness mechanism rather than the change of method. It is possible that multiple imputation under missing-at-random may have delivered a slightly different result to all-randomised standardisation. Sensitivity analyses are important enough that we regard coherence between the primary and sensitivity analysis as worthy of consideration.

\section*{Analyses of \textit{GetTested}\label{s:analysis}}
This article focuses on \textit{planning} a statistical analysis but it is nonetheless instructive to consider some of the issues discussed when different approaches are used. This helps to illustrate what may happen and prompts us to reflect on how we might plan.

As described previously, our analysis of the \textit{GetTested} trial is for two outcome measures: \textit{any~test}, which occurred in $35\%$, and \textit{any~diagnosis}, which occurred in $1.6\%$\cite{wilson17}. For both outcomes, two summary measures were of interest to the investigators: the risk ratio and the risk difference. Recall that there are five categorical covariates to adjust for, which in the direct adjustment model use $18$ parameters in addition to the intercept and randomised arm. Table~\ref{t:gettested} gives the estimated treatment effect and standard error from various analyses estimating each measure on the two outcomes. For the \textit{Any test} and \textit{Any diagnosis} outcomes, there were $612$ and $27$ events respectively from $1,739$ observed outcomes ($324$ had missing outcomes). Table~\ref{t:gettested} presents results for the complete-cases population. Results targeting the all-randomised estimand are presented and discussed in appendix~2. Covariate data were fully observed.

The results presented are intended to compare adjustment methods for the same outcome and summary measure. For the risk ratio, two direct adjustment methods are used: a log-binomial model and a poisson model. In the analyses presented, the methods used included `main' effects of covariates only. Of course, interactions between covariates could have been included for any method, and interactions between covariates and randomised arm could have been included for standardisation. Given the low number of diagnosis events -- both anticipated and observed -- including these interactions would have been inadvisable for that outcome. For these illustrative analyses, missing outcome data were assumed to be missing-at-random given covariates. 

Three of the 14 analyses in table~\ref{t:gettested} failed to produce any sensible estimate. Two of these instances were due to the use of an identity-link-function binomial model to estimate a risk difference, leading to non-convergence, which happened for both outcomes. This emphasises the point that it would have been unwise to plan this as the adjustment model. The last was when directly estimating the log risk ratio for any test using a poisson model. It did produce an estimate, which was 541, which indicated separation of outcome and a clearly untrustworthy estimate.

Of the methods that did converge, the estimated treatment effects and standard errors tended to be similar across methods. The most notable difference in estimates is the complete-cases log~risk ratio for \textit{any diagnosis}, where the direct and standardisation analyses estimated a larger value than the IPTW analysis. This turns out to be due to the covariate \textit{men who have sex with men}, which in the direct model based on poisson regression estimated extremely large risk ratios (around 15). Removing this covariate from the two models resulted in very similar estimates, close to the IPTW result shown.

\section*{Discussion} \label{s:discuss}

We have compared the properties of three broad methods for estimating and adjusted treatment effect: direct adjustment, standardisation and inverse probability of treatment weighting. Our impression is that direct adjustment is the most commonly used approach in clinical trials and that standardisation and IPTW are better appreciated in observational epidemiology and warrant more consideration by trialists. In particular, it is clear that many investigators are interested in summarising the treatment effect as a risk ratio or risk difference. Direct adjustment is notoriously unstable for both measures and so an unadjusted estimate is frequently reported, which will be inefficient. Standardisation and IPTW warrant more use in clinical trials.

Having discussed some properties of the three approaches, we provide table~\ref{t:properties} for reference, listing some of the points to consider for each approach.

We have described the methods in general terms without giving recommendations about specific implementations. In our analysis of \textit{GetTested}, the standardisation estimates were computed from the direct adjustment model (logistic regression including main effect terms for each covariate) but could have been computed from a model including interactions among covariates and of covariates with randomised group. Similarly, we used IPTW on the probability of treatment given covariates, which was estimated using logistic regression with a main effect term for each covariate. 

Tension between methods is typically greatest with binary outcome data, so we chose to focus on this setting. In particular, for continous data analysed using linear models, none of the issues around collapsibility are relevant. Covariate adjustment using analysis of covariance is then typically suitable (direct regression of the outcome on treatment and covariates) unless treatment--covariate interactions are to be modelled, when standardisation would be the appropriate approach.

Binary outcomes are common in non-inferiority trials and the scale of the population summary should match the scale on which the non-inferiority margin is defined. In our experience, the margin is never defined as a conditional odds ratio, suggesting that the use of standardisation or IPTW will be necessary.

For time-to-event outcome data, the issues will depend on the chosen summary measure, as with binary outcome data. In clinical trials, this is frequently the hazard ratio, which is non-collapsible. A further important issue specific to hazard ratios is that even if hazards are proportional on the conditional scale they may not be proportional on the marginal scale, so an adjusted marginal hazard ratio (estimated by standardisation or IPTW) may be an inappropriate summary of the treatment effect, though this is also true of an unadjusted hazard ratio.

It is possible to estimate the covariate-adjusted difference in survival proportion at a time $t$ using standardisation or IPTW, which some trials are beginning to do in statistical analysis plans, or the difference in restricted-mean survival times\cite{royston13}. A variance estimator has been derived for standardisation\cite{sjolander16}, though not (to our knowledge) for IPTW.

Improved power is frequently the main motivation for covariate adjustment, so a natural question arises regarding how to account for this improved power in sample size calculations. The increased power through covariate adjustment depends on the unknown prognostic value of covariates. If this is assumed to be appreciable but turns out to be modest, the trial will be underpowered. We therefore advise a cautious approach and would generally calculate sample size without accounting for covariates. If sample size calculation were to take account of covariate adjustment, we do not believe this would affect the choice between approaches, given that they have similar efficiency.

Many randomised trials report subgroup analyses, which we have not discussed. This raises some interesting issues and questions:
\begin{enumerate}
	\item When adjustment covariates are correlated with the variable defining the subgroup, the expected precision gains from covariate adjustment will diminish. At the most extreme, if we consider adjusting for the variable that defines the subgroup (where the `adjustment' and `subgroup' variables have perfect correlation), adjustment gains nothing: since $X$ does not vary within subgroup, adjustment cannot improve predictions about $Y$.
	\item Should we choose the same approach to adjustment as for the main analyses? Ideally \textit{yes}, with the caveat that any subgroup analysis is inherently conditional on subgroup membership, so the estimand is conditional-on-subgroup. Further, issues around convergence are to be expected in smaller subgroups.
	\item For IPTW, should we fit weighting models separately within subgroups or overall? We believe that this is an open question but note that direct adjustment and standardisation would adjust for covariates separately within subgroups.
	\item Subgroup analyses should be supported by interaction tests. These are straightforward for direct adjustment and IPTW. For standardisation, one could test for interaction on the scale of the working model, but it is more appropriate to test on the scale of the estimand.
\end{enumerate}

\section*{Conclusions}
We hope that this work stimulates statisticians working in trial teams to think carefully about adjustment methods, particularly by placing the estimand -- which requires clinical investigators' input -- first, followed consideration of the more statistical aspects. None of the three approaches is always best and the choice will depend on the trial context. Standardisation and IPTW are largely unused in trials, but have many advantages which mean they warrant routine consideration.

\section*{Appendix 1: targeting the complete-case and all-randomised populations in the presence of missing data}
The received wisdom is that, when missingness depends only on covariates, a complete-case estimator is unbiased and efficient. There is however a subtlety to this, which impacts on our three adjustment methods: The estimand no longer targets a population with the empirical distribution of covariates \textit{among those randomised} but \textit{among the complete cases}. This is true unless the treatment effect is identical within strata on the chosen summary scale, as we will show.

As with non-collapsibility, the fact that the value of the estimand for all-randomised and complete-cases may differ is not bias but a case of targeting different estimands: the target population attribute of the estimands differ. It is perhaps difficult to contrive an argument for the complete-case estimand.

To demonstrate this point and outline some solutions, we will work through a simple numerical example with covariate-dependent missingness, show that a standard complete-case estimator changes the estimand, and note one possible advantage of standardisation over the other methods in this respect. The numerical example can be regarded as deterministic in the sense that estimators that fail to recover the exact value of the estimand seen in the true data do target a different estimand.

Consider a trial with binary $X$, $Y$ and $Z$, where the data can be represented in a $2 \times 2 \times 2$ table. Table~\ref{t:data}(a) gives the full data from a notional randomised trial in which $2,000$ participants are recruited. Of these, $1,000$ have the covariate $X=1$ and $1,000$ have $X=0$. Treatment $Z$ is stratified by covariate $X$ so that $500$ participants have each combination of values of $X$ and $Z$. Finally, table~\ref{t:data}(c) gives the within-stratum odds ratios, risk ratios and risk differences. Note that we have ensured there is some effect modification by $X$ on each of these summary scales, since we will not in general be able to choose a scale on which there is no effect modification (particularly before seeing the data).

Suppose now that among those with $X=0$ exactly half are a complete case, while among those with $X=1$ \textit{all} are a complete case. The complete cases are depicted in table~\ref{t:data}(b). The probability of being a complete case depends only on the covariate $X$, meaning that the frequencies in the cells of the $2 \times 2$ table for $X=1$ are exactly half what they were in table~\ref{t:data}(a). Table~\ref{t:data}(c) thus correctly represents the true values of the within-stratum summaries from table~\ref{t:data}(b) and \ref{t:data}(a).

As previously, we do not dictate which summary of the treatment effect should be used for the trial. We consider four: a risk ratio, a risk difference, a conditional odds ratio, a marginal odds ratio. In these examples, conditional odds ratios are estimated by logistic regression, while marginal odds ratios, risk ratios and risk differences are estimated in two ways: first by standardisation after logistic regression, and second by IPTW. The standardisation analysis does not include an interaction of $X$ with $Z$ in the logistic regression model, though it could do. Note that when two or more approaches can be used for one summary measure (the marginal odds ratio, risk ratio and risk difference are all estimated using standardisation and IPTW), the estimates produced will be identical (see~\cite{hernan20ci}, chapters 12 and 13).

Table~\ref{t:missing} gives results of analyses for all randomised and for the complete cases. The key point is that the true value of the complete-cases estimand differs from the true value of the all-randomised estimand for all summary measures. This is because there is effect modification by $X$ on each scale and the relative proportion with $X=1$ changes in the complete cases under covariate-dependent missingness, so the relative contribution of each stratum-specific effect to the overall effect changes.

It would seem that there is little to choose between the methods in terms of handling of missing data. However, there are different modifications to the adjustment methods that we can used to target the all-randomised estimand in several cases.

We have until now talked about `complete cases' without specifying whether $X$ or $Y$ is incomplete. With $X$ incomplete, the simple-mean-imputation or missing indicator methods describd by White and Thompson can be used to target the all-randomised estimand for marginal summary measures\cite{WhiteThompson05}. Moving beyond complete cases is imperative here since the incomplete cases have important information in observed $Y$. However, neither method is appropriate if the summary measure of interest is conditional and non-collapsible and data are missing-not-at-random, in which case there is no method to target the all-randomised estimand besides correct modelling of the not-at-random missingness mechanism.

With covariate-dependent missingness and outcome $Y$ incomplete, a simple method can be used to target standardisation to the all-randomised estimand:
\begin{enumerate}
    \item Retain in the data all randomised individuals;
    \item Fit the estimation model for standardisation in the complete cases (possibly including additional or different interactions between $Z$ and covariates);
    \item Standardise over the all-randomised distribution of $X$ (implemented in Stata's \texttt{margins} command with the \texttt{noesample} option).
\end{enumerate}
The result of applying this method to the numerical example gives results identical to the left hand column of table~\ref{t:missing}, indicating that the procedure can target the all-randomised estimand. However, as seen in the analysis of the \textit{GetTested} study (see Appendix~2), this neat `trick' is not fool-proof.

For direct adjustment and IPTW, estimators of the all-randomised estimand are slightly less straightforward, but can be achieved by using multiple imputation with a separate imputation model for each randomised arm or using inverse probability of missingness weighting with the weighting model based on randomised arm and covariates. It would seem most natural to pair direct adjustment with by-arm multiple imputation, because both posit a model for the outcome data. IPTW therefore pairs more naturally with inverse probability of missingness weighting, where neither of the weighting models involve the value of the outcome.

\section*{Appendix 2: Issues with estimation for the all-randomised population in \textit{GetTested}}

We now consider further results from re-analysis of \textit{GetTested}. Table~\ref{t:gettested} considered an estimand for the complete-case population, but here we consider the all-randomised population. To target this population, we retain the assumption that outcomes are missing at random and then use a different method for each adjustment approach.

For direct adjustment, we target the all-randomised population using multiple imputation `by-arm'. Each outcome was multiply imputed separately (not jointly), since they are always observed or missing simultaneously and so there is not auxiliary information. The imputation procedure used a logistic regression model, separately for each arm, to impute outcome, including main effects of covariates. The imputation model for both outcomes was `augmented' to handle separation of outcome given covariates\cite{white10pp}. Ten imputations were used. Note that it is the separate imputation model by-arm that targets the all-randomised population. Simply including randomised arm as a covariate in the imputation model would not achieve this unless there were no treatment--covariate interactions on the scale of the summary measure.

For standardisation, the all-randomised estimand was targeted by standardising over the all-randomised population, as described in appendix~1.

For IPTW, inverse-probability-of-missingness weighting was used. The probability of missingness was estimated using a logistic regression model for missingness including all covariates that are present in the primary analysis, and interactions with randomised arm. Since the two outcomes were always missing simultaneously, estimated missingness probabilities were identical for both outcomes. This approach is straightforward to implement due to the monotone missingness pattern (baseline covariates were fully observed and outcomes were missing on the same individuals).

Additional file~2 contains the Stata code used, which can be inspected to resolve any ambiguity in the description of the three methods. Note that, multiple imputation or inverse-probability-of-missingness weighting could have been used for any of the three adjustment methods.

Table~\ref{t:gettestedar} shows the results from these analyses. The main point to note is that some analyses which returned an estimate for the complete-case population (see table~\ref{t:gettested}) do not for the all-randomised. This turns out to be due to collinearity or perfect prediction, described below. However, this affects the methods in different ways, as we see below.

For direct adjustment, the multiple imputation step suffered from perfect prediction due to collinearity and the imputation model for any~diagnosis had to be augmented\cite{white10pp}. The subsequent estimation of risk differences with an identity-link binomial model failed to converge for both outcomes. For the any~diagnosis risk ratios, the log-link binomial model also failed to converge; the poisson model did return an estimate but we do not regard this as a valid all-randomised estimate because many parameters had to be dropped from the imputation models in order for them to fit. For any test, there was severe collinearity in the imputation model. Even after augmentation, only 916 individuals could be used in the imputation model. This again represents a failure to return an estimate relevant to the all-randomised population.

For standardisation, no estimate was immediately returned for either outcome or summary measure. This was due to collinearity in the logistic regression models predicting outcome, which meant that predictions were not produced for four transgender individuals. By omitting these individuals in the standardisation step, an almost-all-randomised estimate could be obtained for any test but not for any diagnosis (marked with asterisks in table~\ref{t:gettestedar}).

IPTW returned estimates for both summary measures and both outcomes. The main threat to it obtaining an estimate for the all-randomised population is perfect prediction of missingness in the missingness model.

Given the above issues, there is litte to say in terms of comparing the estimated log risk ratios obtained across approaches. When approaches did return an estimate and standard error, they were very similar. It is hard to say what differences can be attributed to: the populations targeted, the specific assumptions about missing data, or the approach to modelling assumptions specific to the approach to covariate adjustment. Further investigation of methods targeting the all-randomised population is ongoing. The ideal solution is of course to avoid missing outcome data as far as possible during the conduct of a trial.


\begin{backmatter}

\section*{Acknowledgements}
We are grateful to Jonathan Bartlett, Richard Riley, Leanne McCabe, Brennan Kahan, Andrew Althouse, Maarten van Smeden and Babak Choodari-Oskooei for discussions relating to this work. Our acknowledgement of these individuals does not imply their endorsement of this article.

\section*{Funding}
TPM, ASW and IRW were funded by the MRC grants MC\_UU\_12023/21, MC\_UU\_12023/29, MC\_UU\_00004/07 and MC\_UU\_00004/09. ASW was also funded by the MRC grant MC\_UU\_12023/22. ASW is an National Institute for Health Research Senior Investigator; as such, the views expressed are those of the authors and not necessarily those of the NHS, the NIHR, or the Department of Health. EJW was supported by the MRC Network of Hubs for Trials Methodology HTMR Award MR/L004933/2/N96 and MRC project grant MR/S01442X/1.

\section*{Abbreviations}
IPTW -- inverse probability of treatment weighting\\
STI -- sexually transmitted infection

\section*{Availability of data and materials}
Text for this section\ldots

\section*{Ethics approval and consent to participate}
N/A

\section*{Competing interests}
Tim Morris consults for Kite Pharma, Inc. Ian White has provided consultancy services or courses to Exelixis, AstraZeneca, GSK and Novartis, for which his employer has received funding. Elizabeth Williamson declares personal income from providing training to AstraZeneca.

\section*{Consent for publication}
N/A

\section*{Authors' contributions}
TPM, ASW, EJW and IRW conceived of the article, planned the work and interpreted the results. TPM drafted the article. All authors have approved the submitted version.



\bibliographystyle{unsrt} 





\section*{Figures}
\begin{figure}[h!]
	\includegraphics[width=.5\textwidth]{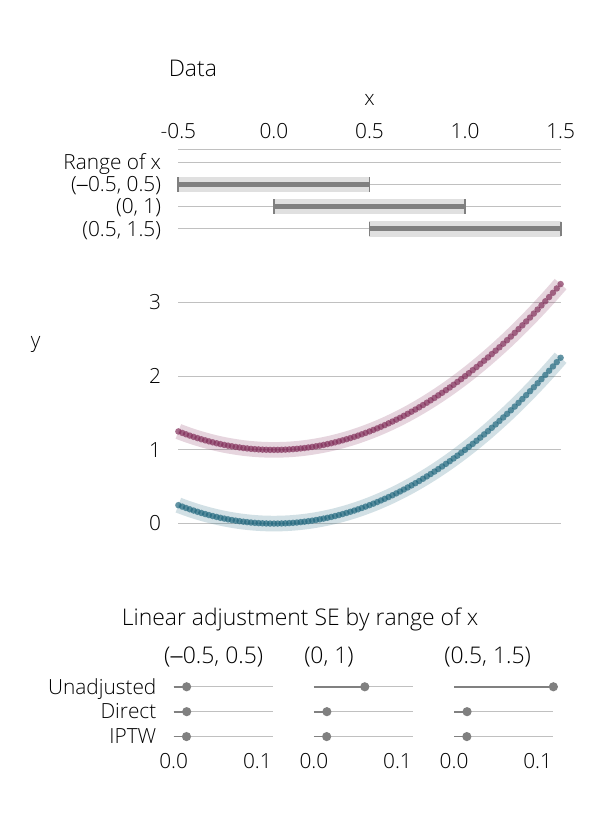}
	\caption{Upper panel: Data from four notional trials where individuals recruited have different distributions of $X$. The two quadratic curves show the data in the two arms. Lower panel: SE after no adjustment and after linear adjustment for each of the three trials. \label{f:quad}} 
\end{figure}



\section*{Tables}
\begin{table}[h!]
    \caption{Some population-level summaries commonly used in clinical trials with binary outcome measures\label{t:psumm}}
	\begin{tabular}{lll}
	\hline
    Outcome type & Summary measures & Collapsible?$^\star$ \\
     & & \\
    \hline
    Continuous & Mean difference & Yes \\
    Binary & Risk difference & Yes \\
     & Risk ratio & Yes \\
     & Odds ratio & No \\
    Time-to-event & Hazard ratio & No \\
     & Restricted mean survival & Yes \\
     & time difference & \\
    \hline
	\end{tabular}\\
    $^\star$Do conditional and marginal summary measures always coincide?
\end{table}

\begin{table}[h!]
\caption{An illustration of non-collapsibility of the odds ratio \label{t:collaps}}
	\begin{tabular}{r|cc|cc|cc}
	\hline
	 & \multicolumn{ 6}{c}{Stratum} \\
	 & \multicolumn{ 2}{c}{A} & \multicolumn{ 2}{c}{B} & \multicolumn{ 2}{c}{Both} \\
	Allocation & Dead & Alive & Dead & Alive & Dead & Alive \\
	 \hline
	 Intervention & 9 & 1 & 5 & 5 & 14 & 6 \\
	 Control & 5 & 5 & 1 & 9 & 6 & 14 \\
	 \hline
	 Odds ratio & \multicolumn{ 2}{c}{9} & \multicolumn{ 2}{c}{9} & \multicolumn{ 2}{c}{5.4} \\
	 \hline
	\end{tabular}
\end{table}

\begin{table}[h!]
\caption{Results of analyses of the \textit{GetTested} trial. All models included main effects only. Link function is canonical unless otherwise specified. The dash symbol~- means model did not converge, except for~* where the log risk ratio estimated as 541, indicating separation for one or more covariates.\label{t:gettested}}
\begin{tabular}{llll|c}
\hline
Outcome       & Summary    & Adjustment      & Model (variable modelled)        & Treatment effect \\
measure       & measure    & method          &                                  & estimate (SE) \\
\hline
Any test      & Risk       & Direct          & Identity-link binomial (outcome) & - \\
(occurred     & difference & Standardisation & Logistic (outcome)               & 0.260 (0.021) \\
in 35\%)      &            & IPTW            & Logistic (treatment)             & 0.262 (0.021) \\
\cline{2-5}
              & Log risk   & Direct          & Poisson, robust SE (outcome)     & -* \\
              & ratio      & Direct          & Log-link binomial (outcome)      & 0.797 (0.075) \\
              &            & Standardisation & Logistic (outcome)               & 0.796 (0.074) \\
              &            & IPTW            & Logistic (treatment)             & 0.806 (0.075) \\
\hline
Any diagnosis & Risk       & Direct          & Identity-link binomial (outcome) & - \\
(occurred     & difference & Standardisation & Logistic (outcome)               & 0.015 (0.006) \\
in 1.6\%)     &            & IPTW            & Logistic (treatment)             & 0.013 (0.006) \\
\cline{2-5}
              & Log risk   & Direct          & Poisson, robust SE (outcome)     & 0.972 (0.433) \\
              & ratio      & Direct          & Log-link binomial (outcome)      & 0.915 (0.412) \\
              &            & Standardisation & Logistic (outcome)               & 0.959 (0.412) \\
              &            & IPTW            & Logistic (treatment)             & 0.855 (0.412) \\
\hline
\end{tabular}
\end{table}

\begin{table}[h!]
\caption{Points to consider on properties of each approach \label{t:properties}}
    \begin{tabular}{ p{3.3cm} | p{3.9cm} | p{3.9cm} | p{3.9cm} }
    \hline
    \textbf{Issue} & \textbf{Direct} & \textbf{Standardisation} & \textbf{Inverse probability} \\
     & \textbf{adjustment} & & \textbf{weighting} \\
    \hline
    Estimand for non-collapsible summary measures & Conditional & Marginal & Marginal \\
    \hline
    For non-collapsible summary measures, true $\beta$ depends on\ldots & Covariates conditioned on in outcome model & In-trial distribution of covariates & In-trial distribution of covariates \\
    \hline
    Misspecification of covariate effects & Loses efficiency vs.\ correctly specified model but expected to gain vs.\ no adjustment. True $\beta$ changes under non-collapsibility & Loses efficiency vs.\ correctly specified model but expected to gain vs.\ no adjustment & Loses efficiency vs.\ correctly specified model but expected to gain vs.\ no adjustment \\
    \hline
    Convergence & Vulnerable & Reasonable (but see \textit{GetTested} experience) & Solid \\
    \hline
    Stratification / minimisation handled by variance estimator & Yes & Yes & Yes \\
    \hline
    Efficiency & Asymptotically optimal & Asymptotically optimal & Asymptotically optimal \\
    \hline
    Standard error calculation & Direct & Delta method & Robust, accounting for estimation of weights via joint estimating equations. Standard error can be biased downwards in small samples\cite{raad20} \\
	\hline
	Treatment--covariate interactions & Can be fitted but does not produce an estimate of an average treatment effect & Naturally handled this and produces an estimate of the average treatment effect & Does not handle \\ 
    \hline
    Handling of missing covariate data in order to target all-randomised population & Missing indicator or single mean imputation (though neither is suitable with non-collapsible population summary measures) & Missing indicator or single mean imputation & Missing indicator or single mean imputation \\
    \hline
    Handling of missing outcome data in order to target all-randomised population & Multiple imputation by-arm (or inverse probability of missingness weighting) & Standardisation to all-randomised rather than complete-case sample; alternatively multiple imputation by-arm or inverse probability of missingness weighting & Inverse probability of missingness weighting (or multiple imputation by-arm ) \\
    \hline
    \end{tabular} \\
\end{table}

\begin{table}[h!]
\caption{a) Full data from a notional randomised trial. b) Complete cases, where all individuals with $X=0$ and half of individuals with X=1 are complete cases. c) True value of summary measure within levels of $X$.
\label{t:data}}
\begin{tabular} {r|cc|cc}
	\multicolumn{ 5}{l}{a) Full data} \\
	\hline
	 & \multicolumn{ 2}{c}{$X=0$} & \multicolumn{ 2}{c}{$X=1$} \\
	 & $Z=0$ & $Z=1$ & $Z=0$ & $Z=1$ \\
	\hline
	 $Y=1$ & 42 & 26 & 180 & 140 \\
	 $Y=0$ & 458 & 474 & 320 & 360 \\
	\hline
	 & 500 & 500 & 500 & 500 \\
	\hline
\end{tabular} ~~~
\begin{tabular} {r|cc|cc}
	\multicolumn{ 5}{l}{b) Complete cases} \\
	\hline
	 & \multicolumn{ 2}{c}{$X=0$} & \multicolumn{ 2}{c}{$X=1$} \\
	 & $Z=0$ & $Z=1$ & $Z=0$ & $Z=1$ \\
	\hline
	 $Y=1$ & 21 & 13 & 180 & 140 \\
	 $Y=0$ & 229 & 237 & 320 & 360 \\
	\hline
	 & 250 & 250 & 500 & 500 \\
	\hline
\end{tabular} \bigskip \\
\begin{tabular} {r|rr}
	\multicolumn{ 3}{l}{c) Value of summary measures within $X$ (both} \\
	\multicolumn{ 3}{l}{in all randomised and among complete cases)} \\
	\hline
	Summary measure & $X=0$ & $X=1$ \\
	\hline
	Odds ratio & $0.598$ & $0.691$ \\
	Risk ratio & $0.619$ & $0.778$ \\
	Risk difference & $-0.032$ & $-0.080$ \\
	\hline
\end{tabular} \medskip \\
\end{table}

\begin{table}[h!]
\caption{Summary of the true value of the estimand in all randomised and among the complete cases under covariate-dependent missingness. \label{t:missing}}
\begin{tabular} {r|rr}
	\hline
    Summary measure  & All-randomised & Complete cases \\
    \hline
	Conditional odds ratio & $0.670$ & $0.679$ \\
	Marginal odds ratio & $0.698$ & $0.700$ \\
    Risk ratio & $0.748$ & $0.761$ \\
	Risk difference & $-0.056$ & $-0.064$ \\
	\hline
\end{tabular} \\
*This method is possible with complete $X$ and incomplete $Y$
\end{table}

\begin{table}[h!]
\caption{Appendix 2: Results of analyses of the \textit{GetTested} trial targeting the all-randomised population. All models included main effects only. Link function is canonical unless otherwise specified. The dash symbol~- means model did not converge, with reasons described in the text of appendix~2.\label{t:gettestedar}}
\begin{tabular}{llll|c}
\hline
Outcome       & Summary    & Adjustment      & Model (variable modelled)        & Treatment effect \\
measure       & measure    & method          &                                  & estimate (SE) \\
\hline
Any test      & Risk       & Direct          & Identity-link binomial (outcome) & - \\
(occurred     & difference & Standardisation & Logistic (outcome)               & 0.258 (0.021)* \\
in 35\%)      &            & IPTW            & Logistic (treatment)             & 0.259 (0.021) \\ \cline{3-5}
\cline{2-5}
              & Log risk   & Direct          & Poisson, robust SE (outcome)     & 0.804 (0.069) \\
              & ratio      & Direct          & Log-link binomial (outcome)      & - \\
              &            & Standardisation & Logistic (outcome)               & 0.799 (0.075)* \\
              &            & IPTW            & Logistic (treatment)             & 0.802 (0.075) \\ \cline{3-5}
\hline
Any diagnosis & Risk       & Direct          & Identity-link binomial (outcome) & - \\
(occurred     & difference & Standardisation & Logistic (outcome)               & - \\
in 1.6\%)     &            & IPTW            & Logistic (treatment)             & 0.013 (0.006) \\ \cline{3-5}
\cline{2-5}
              & Log risk   & Direct          & Poisson, robust SE (outcome)     & - \\ 
              & ratio      & Direct          & Identity-link binomial (outcome) & - \\
              &            & Standardisation & Logistic (outcome)               & - \\
              &            & IPTW            & Logistic (treatment)             & 0.866 (0.414) \\ \cline{3-5}
\hline
\end{tabular} \\
*Almost-all-randomised. Estimate was returned only after omitting four participants affected by collinearity.
\end{table}


\section*{Additional Files}
  \subsection*{Additional file 1 -- Stata code to generate figure~\ref{f:quad}}

  \subsection*{Additional file 2 -- Stata code for the analysis of the \textit{GetTested} trial}
    (Assumes the data file \texttt{journal.pmed.1002479.s001.xls} has been downloaded from https://journals.plos.org/plosmedicine/article?id=10.1371/journal.pmed.1002479\#sec020)

\end{backmatter}
\end{document}